# Building Reusable Software Component For Optimization Check in ABAP Coding

P.Shireesha                          Dr.S.S.V.N.Sharma
Lecturer                             Professor
Dept. of MCA                         Dept. of Informatics
KITS,Warangal,INDIA.                 Kakatiya University,INDIA.
rishapakala@yahoo.co.in              ssvn.sarma@gmail.com

## Abstract

Software component reuse is the software engineering practice of developing new software products from existing    components. A reuse library or component reuse repository organizes stores and manages reusable components. This paper describes how a reusable component is created, how it reuses the function and checking if optimized code is being used in building programs and applications. Finally providing coding guidelines, standards and best practices used for creating reusable components and guidelines and best practices for making configurable and easy to use.

## 1. Introduction

 Reuse allows us to efficiently create reuse of software components improves overall software quality, reduce software costs, and deliver software with fewer defects. Reuse allows us to efficiently create software systems from existing software artifacts rather than building software systems from scratch. Software reuse means reusing the inputs, the processes, and the outputs of previous software development efforts. Effective management of a large set of reusable components requires well-defined structures and processes. Without these, the reuse repository effectively becomes a write-only storage medium. The repository of reusable components is the link between development for reuse, where the components are produced, and development with reuse, where the components are reused.

The initial use of the software reveals any design and implementation faults. These are then fixed, thus reducing the number of failures when the software is reused. Reduced process risk if software exists, there is less uncertainty in the costs of reusing that software than in the costs of development. This is an important factor for project management as it reduces the margin of error in project cost estimation. This is particularly true when relatively large software components such as sub-systems are reused. Effective use of specialists instead of application specialists doing the same work on different projects, these specialists can develop reusable software that encapsulate their knowledge. For example, if menus in a user interfaces are implemented using reusable components, all applications present the same menu formats to users. The use of standard user interfaces improves dependability as users are less likely to make mistakes when presented with a familiar interface.

 Reusable software components are designed to apply the power and benefit of reusable, interchangeable parts from other industries to the field of software construction. Reusable software components can be simple like familiar pushbuttons, textfields, listboxes, scrollbars, dialogboxes. The following figures show reusable software components.

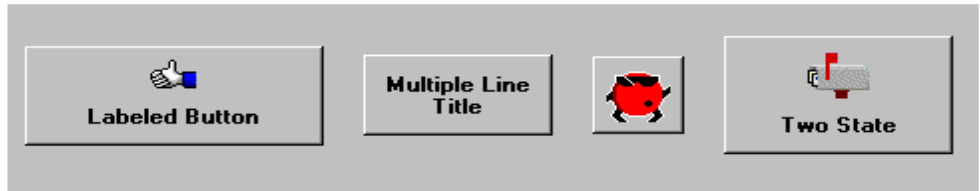

Figure 1. Button beans

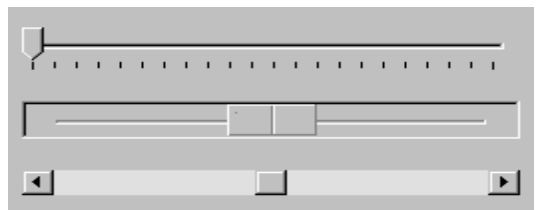

Figure 2.  Slider beans

The software  reusable components need not be limited to source code fragments but also include design structures, specifications, documentation and so on. Based on this reuse can be categorized into four levels[1].
1. Code level components
2. Entire applications
3. Analysis level products
4. Design level products
Code level component reuse occurs most frequently (functional modules, procedures, subroutines, libraries, etc.). However it is criticized for lacking in reuse potential, as the level of abstraction is low for these components. Reusing entire applications often means using off-the-shelf packages, or minimal adaptation of a specialized product applied to a new customer, but it is not always feasible (e.g., real-time software environments). Perhaps the less represented areas are design products, which allow reuse of similar system implementation strategies, and analysis products, which allow reuse of knowledge about real world domains. The analysis level products are possibly the most powerful of all as these allow description and manipulation of real world domains.

The underlying work deals with Code level component reuse. This involves creation of a function module in ABAP and the functionality of the module is to check for optimization of any code written in ABAP. In ABAP, Function modules allow you to encapsulate and reuse global functions in the R/3 System[10].

The rest of this paper is organized as follows. Section 2 is the background giving overview of ABAP and function modules in ABAP. Section 3 provides the proposed work. Section 4 presents the related work and experimentation including Tips and Tricks, function and creating function modules. Next section 5 shows resulting screens and output of function module is portrayed. Finally the paper is concluded along with the future work.

## 2. Background

ABAP is a programming language for developing applications for the SAP R/3 system. ABAP stands for Advanced Business Application Programming Language. ABAP is a programming language for developing applications for the SAP R/3 system. The reusable component is created using function modules, which is one of the modularization techniques in ABAP.

All ABAP programs are modular in structure and made up of processing blocks. There are two kinds of processing blocks, those that are called from outside a program by the ABAP runtime system, and those that can be called by ABAP statements in ABAP programs.

Processing blocks that are called using the ABAP runtime system:

- Event blocks
- Dialog modules

Processing blocks that are called from ABAP programs (also called procedures):

- Subroutines
- Function modules
- Methods

ABAP allows you to **modularize source code** by placing ABAP statements either in local macros or global include programs. This kind of modularization makes ABAP programs easier to read and maintain, as well as avoiding redundancy, increasing the reusability of source code, and encapsulating data. Splitting up ABAP programs into event blocks and dialog modules is designed to help the general flow of the programs.

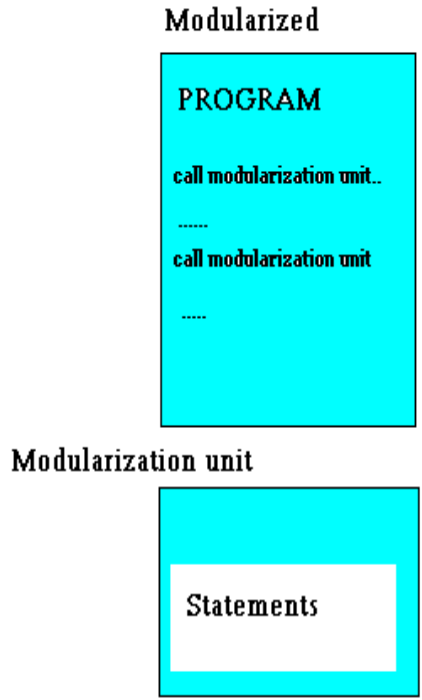

Figure 3. Modularization

## 2.1 Function Modules

Function modules are procedures that are defined in function groups (special ABAP programs with type F) and can be called from any ABAP program. Function groups act as containers for function modules that logically belong together. Function modules allow you to encapsulate and reuse global functions in the R/3 System. They are stored in a central library. Function modules also play an important role in database updates and in remote communications between R/3 Systems or between an R/3 System and a non-SAP system.

Unlike subroutines, we do not define function modules in the source code of your program. Instead, we use the Function Builder. The actual ABAP interface definition remains hidden from the programmer. We can define the input parameters of a function module as optional. We can also assign default values to them. Function modules also support exception handling. This allows you to catch certain errors while the function module is running. We can test function modules without having to include them in a program using the function builder.

## 3. Proposed Work

There is a need to design and use Reusable Components for a product to be developed in ABAP, the programming platform of SAP.

SAP an Enterprise Resources Planning package is used by many organizations for running their business in an optimal way. ABAP/4 is the platform on which the application components are built. It is the language for enhancing SAP and providing plug-ins by programmers.

There is an opportunity for a product based on ABAP
- Which helps organizations in checking if optimized code is being used in building programs and applications
- Which helps the developers in providing coding guidelines and standards
- Which uses best practices used for coding
- Where additions, modifications and deletions of the guidelines and best practices should be configurable and easy to use
- With minimal hard coding to reduce maintenance

To achieve the above, multiple components would be built which are assembled by the application based on the need and order. The main reason for building components is that they can be configured to be called in a specified order and have a state of active or de-active which would be useful in introducing new standards and guidelines without having to tamper with the existing application and its flow.
This would also minimize the redundancies in the code which would help in reducing the number of components that would be impacted due to changes in either the business or technical requirements.

The interfaces of the components would be published for reuse by other components. The input, output and exit criteria for each component would be readily available by which any application can decide if it needs to call a particular component or not. The way the component is implemented is left to the best knowledge of the technical team. With the assurance that the interface of the component would not change once published, would ensure that the applications that are built using these components do not have to be changed due to changes in the components. Any new functionality added to existing components would be readily available to all applications that use these components without any change or downtime.

## 4. Experimentation

To start with the creation of the function module, a lot of analysis has to be performed on the functionality of the reusable component. The main functionality of the module should be to check if a particular code is optimized and if it fails to satisfy the coding standards and guidelines, it should generate a report stating the optimization warning along with example code. For this SAP provides a runtime analysis tool which offers a function 'Tips and Tricks'. In order to make the function module configurable with additions, modifications and deletions of the guidelines and best practices, this function module will form an interface for a number of function modules which uniquely check for one rule or guideline each such that if later one new rule for optimization is included, we can upgrade it with another function module into the main module.

## 4.1 Tips and Tricks

The 'Tips & Tricks' function in the runtime analysis tool contains a series of source code examples intended to illustrate efficient programming. For each problem, it presents two possible solutions, and compares their respective runtimes. The results enable us to see which solution is more efficient. It also provides with features like runtime measurement for the required source code extract. The test screen appears in this format.

This shows two possible solution – 'select…where' and 'select + check' along with runtime which depicts that 'select + check' would require more runtime and hence non-optimized code. If any program uses similar code, on using this function module, it should report this optimization warning and others, if present, in a similar way.

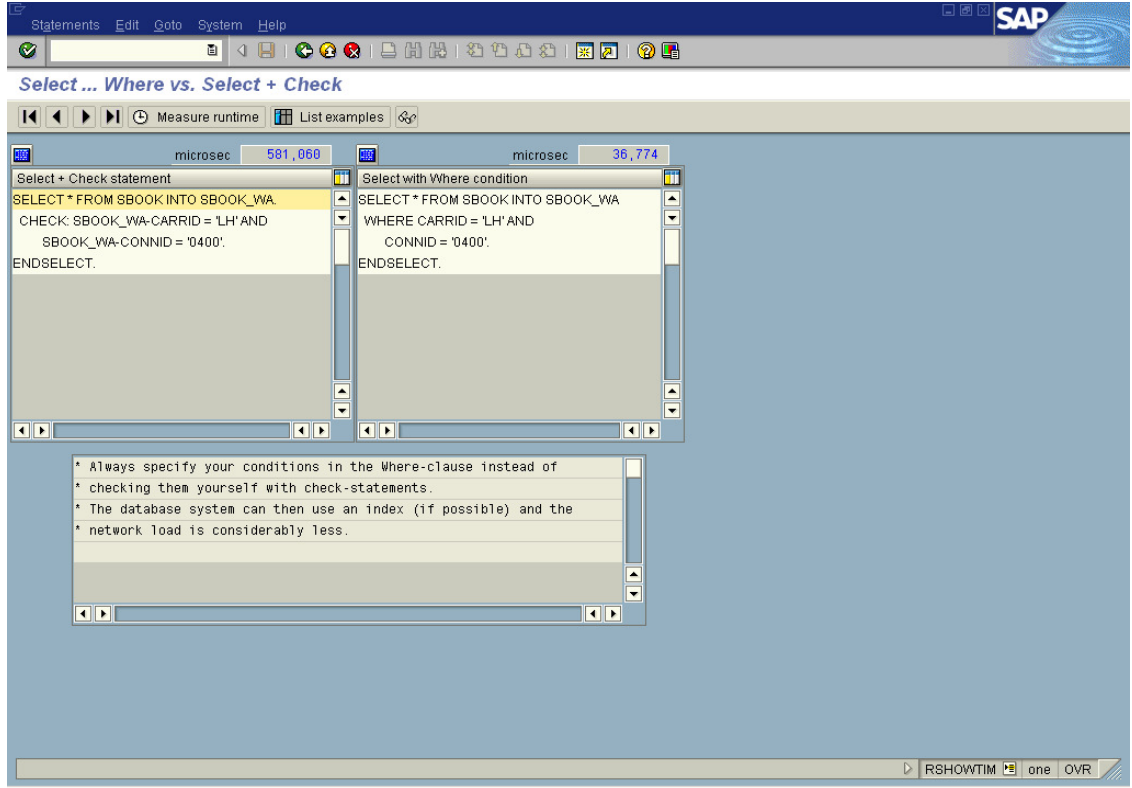

Figure 4. Tips and Tricks

## 4.2 Creating Function modules

We can create function modules and function groups using the Function Builder in the ABAP Workbench. For the proposed work a function module is created which in turn calls function modules that checks for various rules and guidelines and then generates a report accordingly. We can specify the types of interface parameters in function modules in the same way as the parameter interfaces of subroutines. Since function modules can be used anywhere in the system, their interfaces can only contain references to data types that are declared system wide. The function module for optimization check requires an import parameter to read the program name on which the functionality is implemented. Having defined the parameter interface and exceptions, we can write the source code of our function module Function Builder.

The figure 5 shows the Function builder, where the Tab strip shows tabs – Attributes, Import, Export, Changing, Exception and Source code.

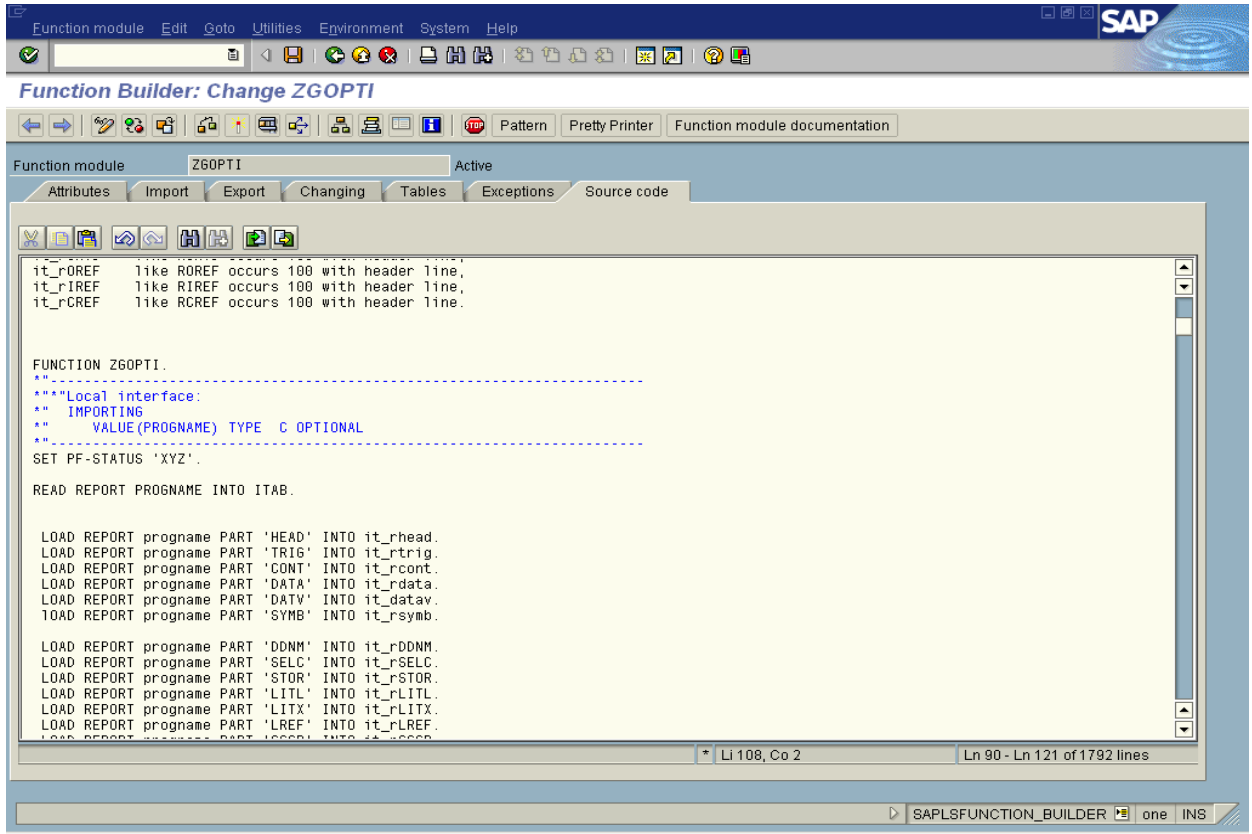

Figure 5. Function Builder

## 5. Results

The main function module forms the interface for a number of function modules that checks for a particular condition (rules and guidelines) each. This interface provides for addition, modification and deletion of new rules and guidelines if it is required.

This main function module can be implemented as a 'CALL FUNCTION' in any ABAP program such that it generates a report which checks for optimization in a particular ABAP coding with the chosen rules and guidelines.

The output of the function module when called in a program with non-optimized code, will be in this format and it is shown in figure 7.

Figure6.  Output of function module

## 6. Conclusion and Future work

In future, there is a plan to add more modules to the component, that provide with new rules and guidelines in ABAP coding.

In this paper, the building of a code level reusable component has been discussed. The purpose of this reusable component as a function module is to check if optimized code is being used in building programs and applications by providing coding guidelines and standards, and also using best practices used for coding.